\documentclass[conference]{IEEEtran}

\def\newblock{\hskip .11em plus.33em minus.07em}

\usepackage{graphicx, booktabs, subfigure, multirow, amssymb, amsmath}
\usepackage[linesnumbered, vlined, ruled, boxed, commentsnumbered]{algorithm2e}

\newtheorem{lemma}{Lemma}[section]
\newtheorem{theorem}{Theorem}

\newtheorem{property}[lemma]{Property}

\begin{document}
%
\title{SECA: Snapshot-based Event Detection for Checking Asynchronous Context Consistency in Ubiquitous Computing}

\author{
    \IEEEauthorblockN{Daqiang Zhang$^{\dag, \ddag}$, Qin Zou$^\S$, Zhiren Sun$^{\dag, \ddag}$}
    \IEEEauthorblockA{$^\dag$School of Computer Science, Nanjing Normal University, Nanjing, 210097, China}
    \IEEEauthorblockA{$^\ddag$Jiangsu Research Center of Information Security \& Confidential Engineering, Nanjing, 210046, China}
    \IEEEauthorblockA{$^\S$School of Remote Sensing and Information Engineering, Wuhan University, Wuhan, 430072, China}
    \IEEEauthorblockA{\{dqzhang, zrsun\}@njnu.edu.cn, qzou@whu.edu.cn}
}

\maketitle

\begin{abstract}
Context-consistency checking is challenging in the dynamic and
uncertain ubiquitous computing environments. This is because
contexts are often noisy owing to unreliable sensing data streams,
inaccurate data measurement, fragile connectivity and resource
constraints. One of the state-of-the-art efforts is CEDA, which
concurrently detects context consistency by exploring the
\emph{happened-before} relation among events. However, CEDA is
seriously limited by several side effects --- centralized detection manner
that easily gets down the checker process, heavy computing complexity
and false negative.

In this paper, we propose SECA: Snapshot-based Event Detection for
Checking Asynchronous Context Consistency in ubiquitous computing.
SECA introduces snapshot-based timestamp to check event relations,
which can detect scenarios where CEDA fails.
Moreover, it simplifies the logical clock instead of adopting the
vector clock, and thus significantly reduces both time and space complexity.
Empirical studies show that SECA outperforms
CEDA in terms of detection accuracy, scalability, and computing
complexity.
\end{abstract}

\IEEEpeerreviewmaketitle

\section{Introduction}\label{sec:intro}


Ubiquitous computing aims to create intelligent environments saturating with computing
and communication capabilities
such that people access ubiquitous applications without knowing underlying technologies. This
intelligence is mainly achieved by context-awareness, which assists
ubiquitous applications in adapting to changeable contexts~\cite{me-mtap}. Contexts
refer to pieces of information that captures the features of ubiquitous computing
environments~\cite{Huang-Percom2009}. The checking of context consistency
is fundamental to ubiquitous computing. For example, a RFID-based system may acquire
two different locations of a user at the same time~\cite{Me-TPDS}. This kind of context consistency
must be checked. However, contexts are often
noisy owing to the unreliable connectivity and resource constraints~\cite{Xu-ICDCS'2008}.
Moreover, contexts frequently keep evolving with user mobility and situations~\cite{me-fgcs, me-ispa2008}.
Therefore, the detection of context consistency is a non-trivial problem.

A variety of schemes for checking context consistency have been
proposed. In~\cite{Wang-Percomw'2004, Bu-BuCLTL06, Bu-QSIC2006}, context
consistency was specified by ontology assertions such that it could
be checked by hidden rules and axioms from ontology.
In~\cite{Xu-2006ICSE, Xu-ICDCS'2008}, context consistency
was modeled by tuples and resolved by \emph{drop-all} and \emph{drop-best}
policies without delineating the checking context consistency.
In~\cite{Xu-2010}, a tree-based checking scheme based on the first-order
logic was reported that checked context consistency by refining consistency trees
using partial context constraints.
However, most existing schemes are seriously limited by two
problems. One is that they are centralized, which incurs their
unscalability in large-scale ubiquitous environments with a huge
number of nodes. The other is that they fall short when counting
in temporal relations among context events, since they implicitly assume
that contexts being checked belong to the same snapshot. But this
assumption cannot be always held in ubiquitous computing environments
that are characterized by asynchronous cooperation and schedule.

To remove the above assumption, CEDA~\cite{Huang-Percom2009} was proposed.
It mapped the context consistency checking into context event detection, and
checked concurrent context events based on the \emph{happened-before} relation.
However, CEDA suffers from three drawbacks, as shown in our previous work~\cite{me-js}.
Firstly, it checked event
detection in a centralized manner, incurring its less
effectiveness in large-scale ubiquitous applications. Secondly, it introduced
false negative because \emph{happened-before} relation
cannot accurately capture all event relations. Finally, CEDA suffered
heavy time and space complexity, which led to its poor
performance.


To this end, we propose in this paper SECA -- snapshot-based event
detection for context consistency checking, which is built on top of
time snapshots and logical clocks. SECA detects context consistency
in a distributed manner, which enables the checking nodes not to
be blocked or become system bottle-necks. It adopts logical clocks
instead of vector clocks to evaluate event relations. To be
scalable, SECA customizes logical clocks by holding the value part.
Thus, it detects event relations that CEDA can and cannot.
Theoretical analysis and extensive experimental results show that
SECA achieves higher detection accuracy than CEDA in a more
scalable manner. The main contributions of this paper are
three-fold.
\begin{itemize}
  \item First, SECA removes the limitation held by CEDA that central-based checking systems are
   easily to get heavy computing load. In contrast, SECA achieves its
   function in a fully-distributed manner, which is highly desirable in large-scale mobile ubiquitous computing.
  \item Second, SECA is capable of detecting false negative scenarios where CEDA fails by introducing the snapshot timestamp.
  \item Finally, SECA respectively reduces CEDA's complexity of time and space for handling an event from $O(n^2)$ to $O(n)$ and from
  $O(n)$ to $O(1)$, where $n$ is the number of nodes in ubiquitous network.
\end{itemize}

The remaining of this paper is structured as follows.
Section~\ref{sec:sca} presents the design of SECA, followed by
theoretical analysis. Section~\ref{sec:experiment} reports our
extensive experimental results. Section~\ref{sec:conclusion}
concludes the paper with a summary and the future work.

\section{SECA: Snapshot-based Event Detection for Checking Asynchronous Context Consistency in Ubiquitous Computing}\label{sec:sca}

Generally, ubiquitous computing environments are modeled as a
loosely-coupled distributed system, where physical entities (e.g.,
objects and users) sense environments, and ubiquitous
infrastructures handle sensed data and deliver services to
ubiquitous applications. In the following, we start with introducing
our system formulation.

\subsection{System Formulation}
Suppose $P_1$, $P_2$, $\ldots$, $P_n$ be $n$ asynchronous processes
in a ubiquitous computing environment, $E_i$ be an event set in
process $P_i$, and $lo$ and $hi$ be the start and end of an event,
respectively. Thus the event $E_i$ is modeled as an interval.
Note that processes communicate with each other
only by means of message-passing, and the communication delay is
finite but unbounded.


In this subsection, we first introduce the concept of snapshot
timestamp and its update policy, and then reshape the
\emph{happened-before} relation with the snapshot
timestamp.

\textbf{Snapshot timestamp.} It refers to an
implementation of the logical clock, where all nodes maintain a logical
clock. In the system of snapshot clocks, the time domain is denoted
as a set of $n--$dimensional and non-negative integer clocks. Each
process $P_i$ maintains a snapshot clock $S_i=\{S_i[k]|k=1,...,n\}$,
where $S_i[k]$ is the $k$th local logical timestamp and describes
the logical time progress at $P_i$. The process $P_i$ updates its
snapshot clock by Rules 1 and 2.
\begin{enumerate}

\item Before sending a message, the process $P_i$ updates its local clock by
    \begin{equation}
    S_i[k] = S_i[k-1] + d (d > 0),
    \end{equation}
    where the default value of $d$ is 1. Then, the process $P_i$ piggybacks a message $m$ with its
    snapshot clock to the remaining nodes in the same environment.

\item When receiving a message $(m, S_j[send])$ from the process $P_j$, the process $P_i$ gets the snapshot timestamp at a $receive$ point as:
    \begin{equation}
    S_i[receive] = max(S_i[k], S_j[send]), (1 \leq k \leq n)
    \end{equation}
\end{enumerate}

\begin{figure}[t]
  \includegraphics[width=1\linewidth]{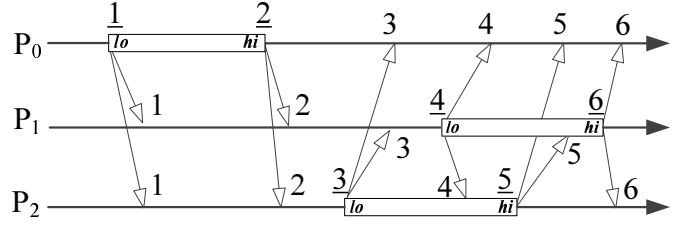}\\
  \caption{An example of snapshot-clock algorithm}\label{fig:snapshot-clock}
\end{figure}

Figure~\ref{fig:snapshot-clock} illustrates the update policies of
our snapshot clock algorithm, where events are represented by the
start and end of intervals --- i.e., $lo$ and $hi$. When the process
$P_0$ would like to send a message, it will automatically increment
the value of its snapshot clock, and then delivers the message to the processes
$P_1$ and $P_2$.

Evidently, by comparing timestamps (i.e., an array of $n-$
elements), the snapshot clock keeps its property of isomorphism. The
relations between timestamp intervals include two ordering relations
represented as '$\leq$' and '$<$', and one concurrent relation
denoted as '$\parallel$'.

\begin{property}\label{property:isomorphism} Given two timestamp intervals $Ip$ and
$Iq$ (a timestamp interval may contain a number of logically continuous timestamps), the isomorphism of the snapshot clock is
given as:
    \begin{eqnarray}
        I_p \leq I_q &\Leftrightarrow& \forall i,i' \, I_p[i] \leq I_q[i'] \nonumber \\
        I_p < I_q &\Leftrightarrow& I_p \leq I_q \textbf{\, and \,} \exists i,i' \, I_p[i] < I_q[i'] \nonumber \\
        I_p \parallel I_q &\Leftrightarrow& \textbf{not \,}(I_p < I_q) \textbf{\, and not \,} (I_q <
        I_p) \nonumber
    \end{eqnarray}
\end{property}

\noindent \textbf{Snapshot-based \emph{happened-before} relation.
 \ } Let '$\rightarrow$' denotes the \emph{happened-before} relation,
the snapshot timestamps based events in a distributed system satisfy
Theorem~\ref{property:snaps-2}.

\begin{theorem}\label{property:snaps-2}
    Given two events $b$ and $c$ with respective timestamp intervals $I_b$ and $I_c$, then:
    \begin{gather}
      b \rightarrow c \Leftrightarrow I_b < I_c \nonumber \\
      b \parallel c \Leftrightarrow I_b \parallel I_c \nonumber
    \end{gather}
\end{theorem}

\begin{IEEEproof}  According to the update policies of snapshot clocks, the
\emph{happened-before} relation is held.
\end{IEEEproof}

Thus, an isomorphism exists between the partially ordered events
produced by a distributed computation and their timestamps. This is
a powerful and interesting property of snapshot clocks. Note that
the \emph{happened-before} relation between these two events is
stated as $b \rightarrow c  \Leftrightarrow  I_b < I_c$.

In order to easily detect concurrent events, we propose an event
concurrence detection mechanism shown as
Theorem~\ref{pro:snap-clock}.

\begin{theorem}\label{pro:snap-clock}
      Given two events $b$ and $c$ in processes $P_i$ and $P_j$. Assume the event $b$ sends a message to the
      event $c$ with its timestamp $I_b.x$, then:

      \begin{center}
      $b \parallel c$  $\Leftrightarrow$  ($I_c.{lo}$ $\leq$ $I_b.x$ $<$ $I_c.{hi}$)\nonumber
      \end{center}
\end{theorem}

\begin{figure}
  \centering
  \includegraphics[width=1\linewidth]{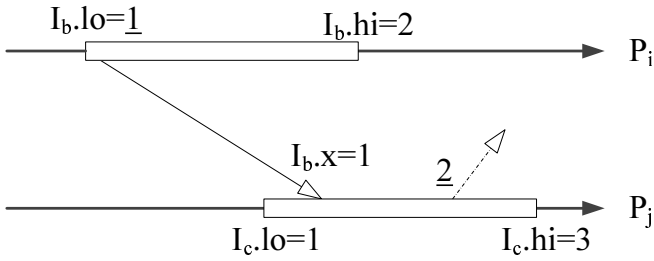}
   \caption{Concurrent events $b$ and $c$, where $I_c.{lo}$ $\leq$ $I_b.x$ $<$ $I_c.{hi}$. }\label{fig:concurrent-event}
\end{figure}

\begin{IEEEproof} There is a message from the event $b$ to the event $c$. According to
the update policy of snapshot clock, the value of $I_b.x$ is not
less than $I_c.lo$. Meanwhile, the message is handled by the event $c$, indicating that
the value of $I_b.x$ must be less than $I_c.hi$. \end{IEEEproof}

\subsection{Snapshot-based Concurrent Event Detection}
In this paper, we propose a SECA scheme --- snapshot-based event
detection for checking asynchronous context consistency in
ubiquitous computing. SECA is built on top of the snapshot
timestamps, and enables all nodes to detect concurrent context
consistency events without central control or a centralized
hierarchy. In SECA, the basis of context consistency detection is
Theorem~\ref{pro:snap-clock} and \emph{happened-before} relation. Figure~\ref{fig:concurrent-event}
illustrates the fact that the events $b$ and $c$ are concurrent, which is checked by
Theorem~\ref{pro:snap-clock}.

The pseudo-code of SECA scheme is given as
Algorithm~\ref{algorithm:process}, consisting of three parts: event
processing, message processing, and context consistency checking.
The event processing refers to a process that updates its snapshot
clock when an event occurs within its life span. To be specific, the
process updates its snapshot clock, the event queue $EQ$, as well as
interval queue $IQ$ by broadcast (e.g., SECA offers a
$System\_Broadcast$ primitive). When events communicate with
messages, the processes where the events happen meet two types of
message processing
--- sending and receiving. The sender is in charge of updating
the event queue and interval queue (see steps 11-14).
Correspondingly, the receiving process modifies its snapshot clock
by picking the maximal timestamp value between the snapshot
timestamps of the sender and receiver processes (see steps 15-19).
Note that the actions of senders and receivers are incorporated
together in the pseudo-code. The third part refers to the context
consistency detection. Since elements in $EE$ implicitly satisfy
Theorem~\ref{pro:snap-clock}, we output the event pairs simply by a
validation check.

\begin{algorithm}


\KwIn{$P=\{P_1, \ldots, P_i, \ldots, P_n\}$, a process set in a
ubiquitous;
system\\
    \qquad \quad $EQ[]$, an event queue;\\
    \qquad \quad $IQ[]$, an interval queue;\\
    \qquad \quad $EE[]$, pairs of events which have communication with each other;  \\
    \qquad \quad $S[]$,  a list of snapshot timestamps;
}

\KwOut{A set of concurrent events $C = \{<e_x, e_y>\}$}

\Begin{
    \tcc{\small{When an event $e$ occurs at $P_i$}}
    \If{$P_i \overset{occur}{\longleftarrow} e$}
    {
        \tcc{\small{suppose it occurs at timestamp k with id $e_{id}^i$}}
        $S_i$[k]=$S_i$[k-1]+1; \\
        $EQ[i]$.push($e_{id}^i$, $S_i$[k]);\\
        $IQ[i]$.push($e_{id}^i$, $lo$=$S_i$[k], $hi$=$S_i$[k]+1);\\

        System\_Broadcast($P_i$, $S_i$);
    }

    \tcc{\small{Upon the process $P_i$ receives a message from the process $P_j$}}
    \If{$P_i \overset{msg}{\longleftarrow} P_j$}
    {
        ($e_{id}^{j}$, $S_j$[k]) = $EQ[j]$.getTop();\\
        $S_j$[k+1] = $S_j$[k]+1;\\
        $EQ[j]$.push($e_{id}^{j}$, $S_j$[k+1]);\\
        $IQ[j]$.current = ($e_{id}^j$, $lo$, max($hi$, $S_j$[k+1]+1);\\
        $S_i[receive]$ = max($S_i$, $S_j$[k+1]);\\
        \If{$e_{id}^j$ is received by $e_{id}^i$}
        {
            ($e_{id}^i$, $lo$, $hi$) = $IQ[j]$.pop();\\
            $IQ[i]$.push($e_{id}^i$, $lo$, max($hi$, $S_j$[k+1]);\\
            EE.push($e_{id}^i$, $e_{id}^j$);
        }
    }

    \BlankLine
    \tcc{\small{Context consistency detection}}
        \While{(!EE.IsNullOrEmpty())}{
            $<e_{id}^i, e_{id}^j>$ = EE.pop();\\
            \If{(IsValid($e_{id}^i$))
            $\&\&$ (IsValid($e_{id}^j$))}{
            $C$.push($e_{id}^i$, $e_{id}^j$);
            }
    \tcc{\small{Output concurrent events}}
            Unique($C$);
        }

    \BlankLine
} \caption{SECA checks context consistency in normal processes
}\label{algorithm:process}
\end{algorithm}

\subsection{Discussions}
Thus far we have presented the design of SECA. However,
does SECA scheme solve false negative caused in CEDA scheme? Can SECA scheme
detect context consistency accurately in ubiquitous computing
environments? We investigate these issues by theoretical analysis in
this section. Specially, we will study the false negative, complexity and implementation
manner of the proposed scheme. Moreover, in the following Section~\ref{sec:experiment}, we
further evaluate SECA by extensive experiments.

\subsubsection{False Negative in happened-before-based Context Consistency
Detection} 
\begin{figure}[t]
  \centering
  \includegraphics[width=1\linewidth]{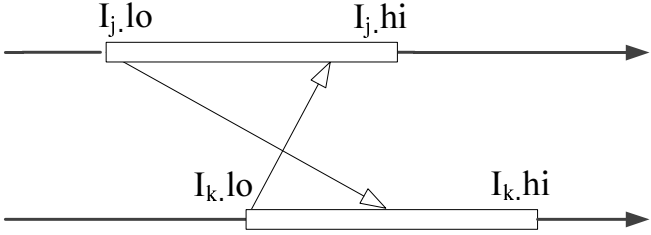}\\
  \caption{Overlapping intervals that can be detected based on \emph{happened-before} relation in CEDA scheme}\label{fig:ceda-detection}
\end{figure}

Given $n$ intervals $I_1$, $I_2$, $\ldots$, $I_n$, CEDA
checks concurrent context consistency events by Eq.~\ref{eq:ceda},
which is built on top of \emph{happened-before} relation. The case
of interval overlaps which is characterized by concurrent events
shown as Fig.~\ref{fig:ceda-detection}. However, for some
concurrent events whose intervals are overlapped, the CEDA scheme fails to
detect them, which is notorious for \textbf{false negative}
phenomena. This is because Eq.~\ref{eq:ceda} cannot detect these
overlapping intervals although they are mutual across.
\begin{eqnarray}\label{eq:ceda}
  (I_j.{lo} \rightarrow I_k.{hi}) \wedge (I_k.{lo} \rightarrow I_j.{hi}), \forall 1 \leq j \neq k
  <n.
\end{eqnarray}

\begin{figure*}[htb!p]
    \centering
  \subfigure[]{
    \label{fig:false-negative-a}
    \begin{minipage}[b]{0.32\textwidth}
      \centering
      \includegraphics[width=0.93\linewidth]{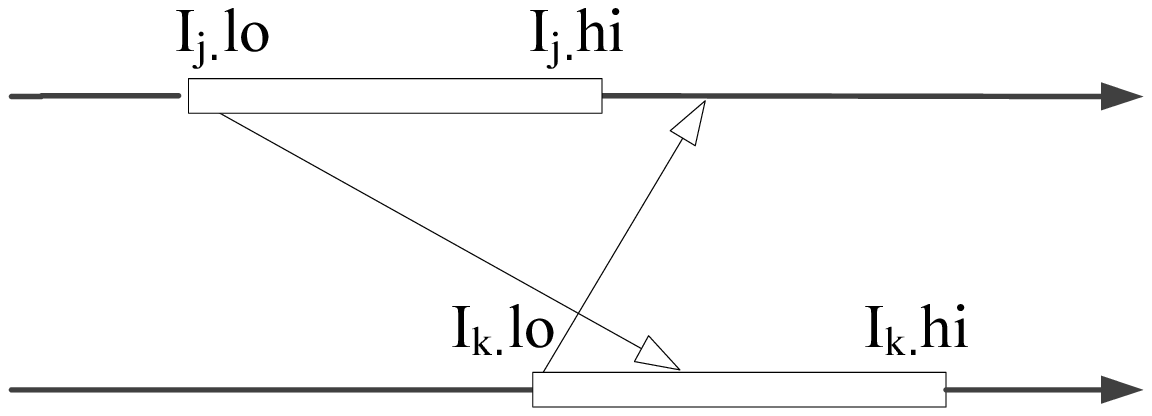}
    \end{minipage}}%
  \subfigure[]{
    \label{fig:false-negative-b}
    \begin{minipage}[b]{0.32\textwidth}
      \centering
      \includegraphics[width=0.93\linewidth]{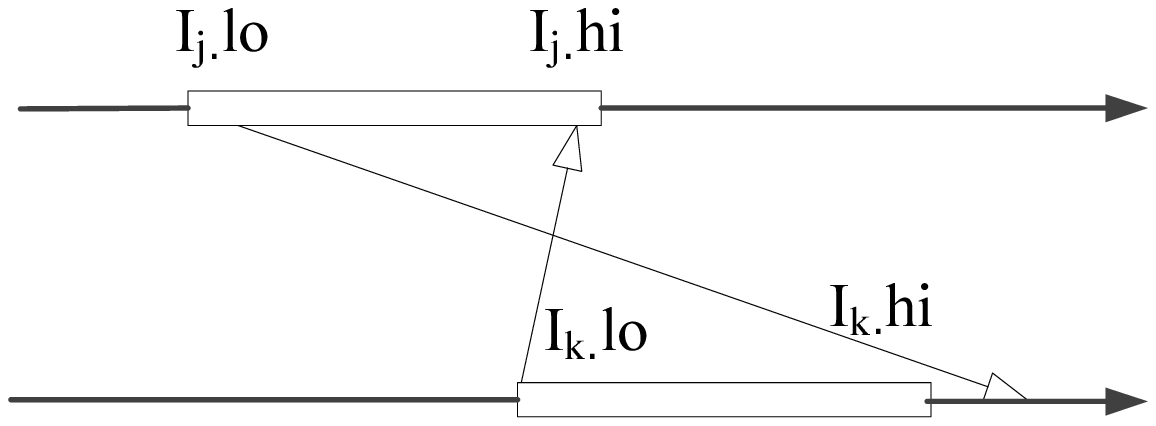}
    \end{minipage}}
     \subfigure[]{
    \label{fig:false-negative-c}
    \begin{minipage}[b]{0.32\textwidth}
      \centering
      \includegraphics[width=0.93\linewidth]{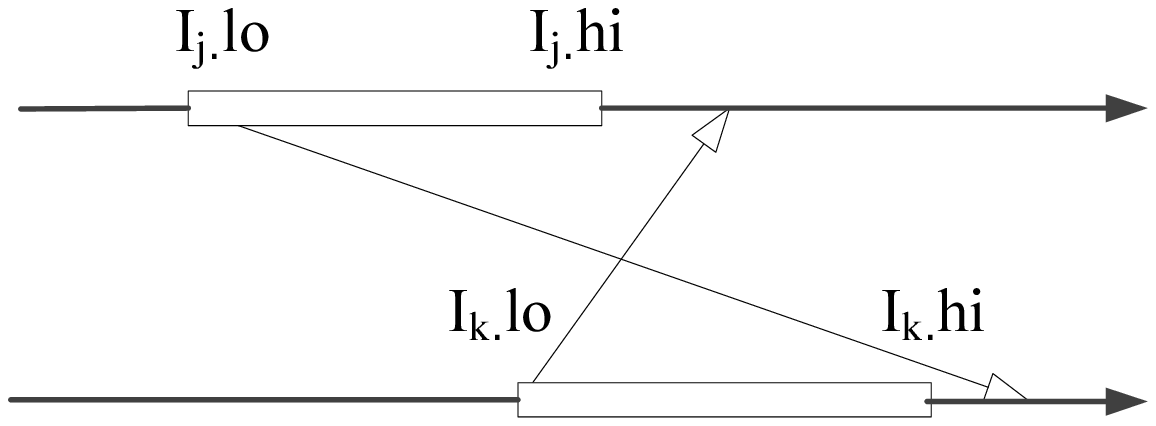}
    \end{minipage}}
  \caption{Three kinds of false negative scenarios caused by \emph{happened-before} relation in CEDA scheme.
  Concurrent events in scenarios (a) and (b) can be accurately detected by SECA scheme.}
  \label{fig:ceda-false-negative}
\end{figure*}

Figure~\ref{fig:ceda-false-negative} illustrates three false
negative scenarios where CEDA scheme fails to check context
consistency correctly. In
Figs.~\ref{fig:false-negative-a},~\ref{fig:false-negative-c}
and~\ref{fig:false-negative-c}, two events in two respective processes satisfy
$(I_j.{lo}\rightarrow I_k.{hi}) \wedge (I_k.{lo}\nrightarrow
I_j.{hi})$, $(I_j.{lo}\nrightarrow I_k.{hi}) \wedge
(I_k.{lo}\rightarrow I_j.{hi})$, and $(I_j.{lo}\nrightarrow
I_k.{hi}) \wedge (I_k.{lo}\nrightarrow I_j.{hi})$, respectively.
These two events concurrently take place, but Eq.~\ref{eq:ceda}
are blind of them. On the contrary, SECA is capable of
successfully identifying these concurrent context events. As for
Figs.~\ref{fig:false-negative-a} and~\ref{fig:false-negative-b}, SECA
compares the message timestamp of senders with the $lo$ and $hi$ of
the receivers and then locates the concurrency. Note that concurrency
in Fig.~\ref{fig:false-negative-c} is challenging to detect. This
kind of concurrency is mainly caused by message delay. Owing to the space limitation,
we only report the experimental results about how message delay affects the SECA performance, and
omit the part of theoretical analysis.



\subsubsection{Complexity} Taking a panoramic view of the SECA scheme, it is easy to
find that SECA does not rely on central control to check context
consistency. All processes involved in a ubiquitous system are equal
and check context consistency by snapshot clocks. Every process
requires $O(1)$ space complexity to maintain snapshot timestamps,
and $O(n)$ time complexity for every context consistency event
detection. Considering that many ubiquitous devices are with limited computing and
communication capabilities, SECA is highly desirable with respect to efficiency and
scalability in large-scale mobile ubiquitous environments.

To further evaluate the time and space complexity of the proposed
scheme, we have implemented the detection schemes by physical
clocks, vector clocks and snapshot clocks, labeled as PCA, CEDA and
SCA schemes, correspondingly. Table~\ref{tbl:theorectical-comparison} compares
the PCA, CEDA and SCA in terms of clock synchronization, handling
the occurrence of an event, detecting overlapped intervals and
concurrent events, and false negative. By comparison, SCA
significantly alleviates the time and space complexity concerning event
processing and context consistency checking. Meanwhile, SCA also
cuts off a half possibility of false negative generated in CEDA
scheme.

\begin{table}[tp]\caption{Comparison of PCA, CEDA and SCA schemes with respect to checking
context consistency events} \label{tbl:theorectical-comparison}
\footnotesize
\begin{tabular}{llll}
  \toprule
  Items                    &PCA                &CEDA           &SCA \\  \midrule
  Synchronization          &$\surd$  &$\times$       &$\times$ \\
  An event occurs&$\times$             &$O(n)$         &$O(1)$\\
  Concurrent events    &$O(n)$   &$O(n^2)$      &$O(n)$\\
  False negative           &$\times$      &$|overlap|<\varepsilon$ &$-\varepsilon<overlap<0$ \\
   \bottomrule
\end{tabular}%
\end{table}

\subsubsection{Implementation Manners}
In general, SECA scheme can be achieved in a distributed manner, the same as that of \emph{happened-before}
relation implementation in distributed systems. This kind of implementation is appropriate to PCs and
supercomputers that are equipped with powerful communication and computation capabilities. With respect to
the hand-held and embedded devices, e.g., sensors, RFID, and mobile phones, the proposed scheme can
be reached by agents, e.g., mobile agents for RFID and mobile phones~\cite{me-ksii}.

\section{Experiments}\label{sec:experiment}

We conduct extensive experiments in this section to further evaluate
whether SECA is appropriate to context-aware applications in
asynchronous ubiquitous computing environments. In particular, this
section will evaluate how the detection accuracy of SECA is, and
whether SECA outperforms CEDA regarding detection accuracy and
computation cost.

\subsection{Experiment Setup} A smart building scenario is simulated where users
move around. The duration of users' stay in an office follows the
exponential distribution. In view of that user location is regarded
as the most important type of contexts in asynchronous ubiquitous
computing
environments~\cite{Xu-ICDCS'2008, landmarc, Active-Badge},
the user location is our focus. The holistic study environment is
equipped with RFID devices and every user carries a RFID tag such
that the location context is collected timely. The
RFID data concerning user location is generated with controlled
error rates of 10\%, 20\%, 30\%, 40\% and 50\% by leveraging on the
mechanisms provided in the existing
literature~\cite{Xu-2010, Rao-DCM}. A constraint is implanted into the location context
that \textbf{\emph{a user cannot have two difference locations at
the same time}}.



\begin{figure*}[htb]
    \centering
  \subfigure[Overall performance by increasing node participants]{
    \label{fig:false-negative-a}
    \begin{minipage}[b]{0.5\textwidth}
      \centering
      \includegraphics[width=1\linewidth]{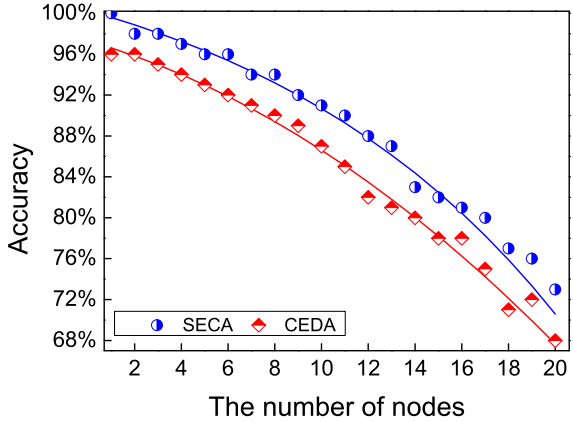}
    \end{minipage}}%
  \subfigure[Detection accuracy with varying message delays]{
    \label{fig:false-negative-b}
    \begin{minipage}[b]{0.5\textwidth}
      \centering
      \includegraphics[width=1\linewidth]{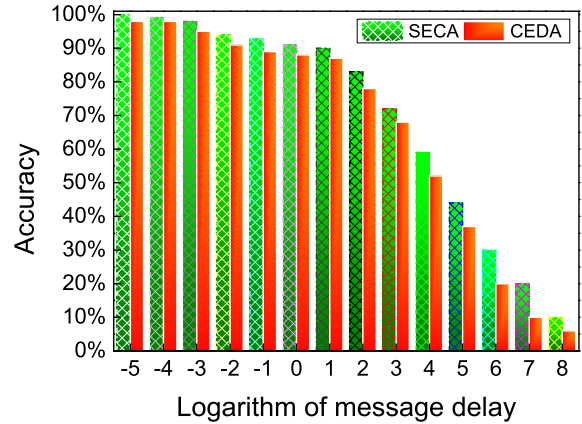}
    \end{minipage}}
  \caption{Performance evaluation of SECA and CEDA schemes.
  }
  \label{fig:performance}
\end{figure*}

\subsection{Overall performance}
A series of experiments is designed to check the
detection accuracy of SECA and whether it performs better than CEDA.
Given that the experiments shed light on detecting concurrent
events of user locations, we limit the number of nodes attending for
the same contexts from 2 to 20. Every node runs two detection
process instances. Every event has a random life span from 20
milliseconds to 50 milliseconds and every message suffers a random
delay between 0.25 to 8 seconds. All experimental results are gained
by PC equipped with Windows Enterprise 7 (32-bit), CPU 1.67GHz, and RAM 2GB. The
following experiments share the same settings without explicit
declaration.

Figure~\ref{fig:performance}(a) illustrates the performance results
with tuning the number of nodes from 2 to 20. Both CEDA and SECA
schemes achieve a high level of detection accuracy of context
consistency events, showing a slightly downward trend. This
indicates that vector clocks and \emph{happened-before} relation is
efficient for detecting concurrent context consistency events.
Meanwhile, SECA scheme acquires a higher level of accuracy than CEDA. This is
because SECA correctly solves part of cases where CEDA gets false negative errors.

\subsection{Detection performance with varying message delays}

Several experiments are conducted to investigate how the message
delay affects the concurrent event detection of the proposed scheme.

As shown in Fig.~\ref{fig:performance}(b), detection accuracy of
both SECA and CEDA schemes reduce their accuracy as the increase of message
delay. In all experiments, SECA achieves a higher level of accuracy
than CEDA owing to its snapshot-based timestamp checking mechanism.
Especially, when the logarithm of the message delay is between -2
and 3, SECA gets a better detection accuracy with less communication
overheads. Taking into account the scale of ubiquitous network, we
hereby set the value of message delay as 0.25 to 8 milliseconds.

%
%

\section{Conclusion}\label{sec:conclusion}

In this paper, we have studied concurrent event detection for
checking context consistency in asynchronous ubiquitous
environments. We have proposed the snapshot timestamp and based on
it we have put forward the SECA scheme, which reduces the time complexity of CEDA
from $O(n^2)$ to $O(n)$, and the space complexity for handling an event from $O(n)$ to $O(1)$, where
$n$ is the scale of ubiquitous network. Extensive experimental
results show that SECA is desirable in context-aware applications
and outperforms CEDA regarding concurrent event detection
accuracy, and robustness on message delay and event duration.

Currently, SECA scheme could be further improved in the following
perspectives. Firstly, we need to investigate how SECA performs in
large-scale ubiquitous computing environments with over ten thousands of
participants. Secondly, we will study whether and how SECA copes
with the dynamic changes of processes involved in the concurrent
event detection. Finally, we plan to evaluate SECA
in various scenarios with more types of contexts and consistency
constraints.

\section*{Acknowledgement}
This work is supported by the National Natural Science Foundation of
China (Grant Nos. 61103185, 61100178, 61003247 and 61073118), the Start-up
Foundation of Nanjing Normal University (Grant No. 2011119XGQ0072), and
Natural Science Foundation of the Higher Education Institutions of Jiangsu
Province, China (Grant No. 11KJB520009). This work is also supported by Major
Program of National Natural Science Foundation of Jiangsu Province (Grant No.
BK211005).  This research was also partially supported by the National Research
Foundation of Korea (NRF) grant funded by the Korea government (MEST) (Grant
No. 2011-0009454).

We would like to thank Ms. Tianyi Zhan for her initial implementation. We
also thank Nicole Kowh for her discussions and proofread.


\IEEEtriggercmd{\enlargethispage{-5in}}


%

%
%



\end{document}